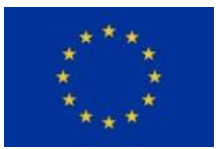 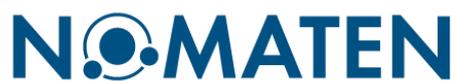

This work was carried out in whole or in part within the framework of the NOMATEN Centre of Excellence, supported from the European Union Horizon 2020 research and innovation program (Grant Agreement No. 857470) and from the European Regional Development Fund via the Foundation for Polish Science International Research Agenda PLUS program (Grant No. MAB PLUS/2018/8), and the Ministry of Science and Higher Education's initiative "Support for the Activities of Centers of Excellence Established in Poland under the Horizon 2020 Program" (agreement no. MEiN/2023/DIR/3795).

# CoNi-MOF laccase-like nanozymes prepared by dielectric barrier discharge plasma for treatment of antibiotic pollution

Chao Liu[a,b], Yi Cao[a,b], Qi Xia[a], Amil Aligayev[a,b,c], Qing Huang[a,b,*]

[a] CAS Key Laboratory of High Magnetic Field and Iron Beam Physical Biology, Institute of Intelligent Machines, Hefei Institutes of Physical Science, Chinese Academy of Sciences, Hefei 230031, China
[b] Science Island Branch of Graduate School, University of Science and Technology of China, Hefei 230026, China
[c] NOMATEN Centre of Excellence, National Center for Nuclear Research, 05-400 Swierk/Otwock, Poland

Corresponding author: Prof. Qing Huang (Email: huangq@ipp.ac.cn)

**Abstract:** Laccase is a natural green catalyst and utilized in pollution treatment. Nevertheless, its practical application is constrained by limitations including high cost, poor stability, and difficulties in recovery. Herein, with inspiration from catalytic mechanism of natural laccase, we designed and prepared a bimetallic metal-organic framework, namely, CoNi-MOF, using low-temperature plasma (LTP) technology. We employed dielectric barrier discharge (DBD) plasma to prepare CoNi-MOF, and by precisely modulating the $N_2/O_2$ gas ratio, we could modulate the distribution concentration of oxygen vacancies in CoNi-MOF. Experimental investigations and density functional theory (DFT) calculations elucidated that the critical role of the oxygen vacancies in enhancing the laccase-like activity, which promoted the activation of molecular oxygen ($O_2$) for generation of reactive oxygen species (ROS). Compared to natural laccase, CoNi-MOF exhibited superior catalytic performance in the degradation of antibiotic tetracycline (TC), along with enhanced resistance to harsh environmental conditions, improved stability, and low biotoxicity. Notably, aeration increased the dissolved oxygen (DO) content, further improving the TC degradation efficiency. As such, this study not only proposes a facile and efficient low-temperature plasma technology for synthesizing high-performance laccase- like nanozymes but also provides a promising and environmentally friendly strategy for the remediation of antibiotic contamination in the environment.

**Environmental implication**

Tetracyclines are widely used broad-spectrum antibiotics that can pose significant ecological and biological risks. Consequently, it is crucial to efficiently remove high concentrations of tetracyclines from pharmaceutical wastewater, medical waste, and livestock manure. Taking inspiration from the active site of the natural green catalyst, laccase, we designed and prepared a bimetallic metal-organic framework, namely, CoNi-MOF, using low-temperature plasma (LTP) technology. The CoNi-MOF with excellent laccase-like activity could effectively catalyze the reactions leading the degradation of tetracycline with high-concentrations, along with enhanced resistance to harsh environmental conditions, improved stability, and minimized biotoxicity.

**Introduction**

Emerging contaminants (ECs), including antibiotics, steroids, plasticizers, and antimicrobial agents, pose significant threats to human and animal health due to their neurotoxicity, immunotoxicity and carcinogenicity [1]. To address these concerns, researchers have developed and implemented diverse remediation strategies targeting the elimination of EPs from various environmental matrices, such as physical adsorption, advanced oxidation processes, and photocatalytic degradation [2-4]. However, these traditional physical/chemical methods often suffer from low efficiency, high energy consumption, and the potential generation of toxic by-products. In contrast, enzyme-mediated bio-transformation, an emerging biotechnology, has gained attention as one of the most promising approaches for ECs removal. This method not only transforms organic pollutants into less toxic substances but also offers advantages such as low cost and environmental friendliness [5]. Tetracycline (TC), a broad-spectrum antibiotic widely used in aquaculture, modern medicine, and the food industry, has become a significant environmental pollutant due to its extensive misuse. Its accumulation in the environment and subsequent entry into the human body through the food chain can lead to severe health issues [6,7]. Therefore, the effective removal of antibiotics such as TC from the environment is of paramount importance.

Laccase, a multi-copper oxidase, is a green catalyst that utilizes oxygen as an electron acceptor, with water as the reduction product [8]. It exhibits broad substrate specificity, enabling the catalysis of various phenolic compounds and the effective degradation of water pollutants. As a result, laccase has been widely employed in treating industrial wastewater and has found applications in industries such as papermaking, food processing, and pharmaceuticals. However, the practical application of natural laccase is hindered by several limitations, including high cost, poor stability, and difficulties in recovery [9,10]. Although various methods have been developed to mimic the activity of natural laccase, these approaches often involve complex procedures, high costs, and limited enzyme yields.

Nanozymes, being the nanomaterials designed to mimic natural enzymes, have emerged as a promising alternative [11]. They offer advantages such as cost-effectiveness, robustness, recyclability, biocompatibility, and tunable physicochemical properties, making them highly suitable for environmental remediation [12]. Efficient nanozymes are characterized by their ability to generate reactive oxygen species (ROS) and their excellent dispersibility in solutions, which are critical for pollutant degradation. Despite the potential of laccase-like nanozymes in environmental applications, most reported nanozymes still fall short of the high catalytic performance required for effective pollutant degradation. Challenges include the lengthy and low-yield preparation processes, which result in high costs, as well as the need to further enhance catalytic activity and treatment efficiency [13]. Therefore, developing more efficient and scalable preparation methods, particularly those that are low-cost and capable of large-scale production, remains a significant challenge. Metal-organic frameworks (MOFs), porous nanomaterials composed of metal nodes and organic ligands, have gained attention for their high surface area, porosity, and tunable structures. They have been widely applied in pollutant treatment [14]. However, most reported laccase-like MOFs are copper-based and suffer from time-consuming and low-yield preparation methods. For instance, Li et al. reported a hydrothermal method to synthesize a laccase-mimicking nanozyme (Cu-ATZ) with a high $Cu^+$ ratio for tetracycline removal [15]. Currently, researchers are still striving to develop alternative and scalable methods for preparation of high-efficiency nanozymes, and especially, it is of great interest to establish new effective methods for preparation of MOF-based nanozymes.

Fortunately, we recently demonstrated that low-temperature plasma (LTP) could serve as an effective, rapid, and environmentally friendly method for the large-scale production of MOF-based nanozymes [16]. LTP, induced by high-voltage electron discharge at ambient temperature, creates a non-equilibrium system of electrons and ions, and the excited state of plasma makes it a powerful tool for inducing rich chemical reactions, which are essential for modifying and preparing new materials. For example, Chen et al. used high-energy ions (e.g., $Ar^+$, $N_2^+$, $H_2O^+$) in plasma to etch metal oxide surfaces, selectively removing surface oxygen and creating oxygen vacancies, thereby exposing more active sites for reactions [17]. In our recent work, we employed the technique of LTP in the nanomaterial preparation, where we imposed plasma jet into the solution, and successfully fabricated Cu-MOF nanozymes that could be utilized as sensors for detecting bioactive components in food [16].

In this work, we employed a common LTP form of plasma, namely, dielectric barrier discharge (DBD) plasma to prepare CoNi-MOF laccase-like nanozymes. With the assistance of abundant reactive species generated by DBD, $Ni^{2+}$ and $Co^{2+}$ rapidly coordinated with imidazole ligands to form CoNi-MOF nanoflowers. Specifically, the energy from DBD promoted the coordination of metal ions with ligands, while various reactive species generated a lot of oxygen vacancies on the CoNi-MOF surface. The entire preparation process could be completed within one minute. This unique CoNi-MOF, featured with cyclic electron pairs and abundant reactive sites, exhibited excellent laccase-like activity. Both experimental and theoretical results demonstrated that the presence of oxygen vacancies in CoNi-MOF significantly influenced its laccase-like activity. As a practical application example, we applied the CoNi-MOF nanozyme for the treatment of tetracycline, showcasing its potential for antibiotic removal. Scheme 1 illustrates the MOF-based biomimetic design of natural laccase active sites, the LTP preparation method, and the application in antibiotic pollution treatment. This study not only introduced an innovative and efficient technology for synthesizing high-performance laccase-like nanozymes but also established a promising and eco-friendly strategy for the remediation of antibiotic contamination in the environment.

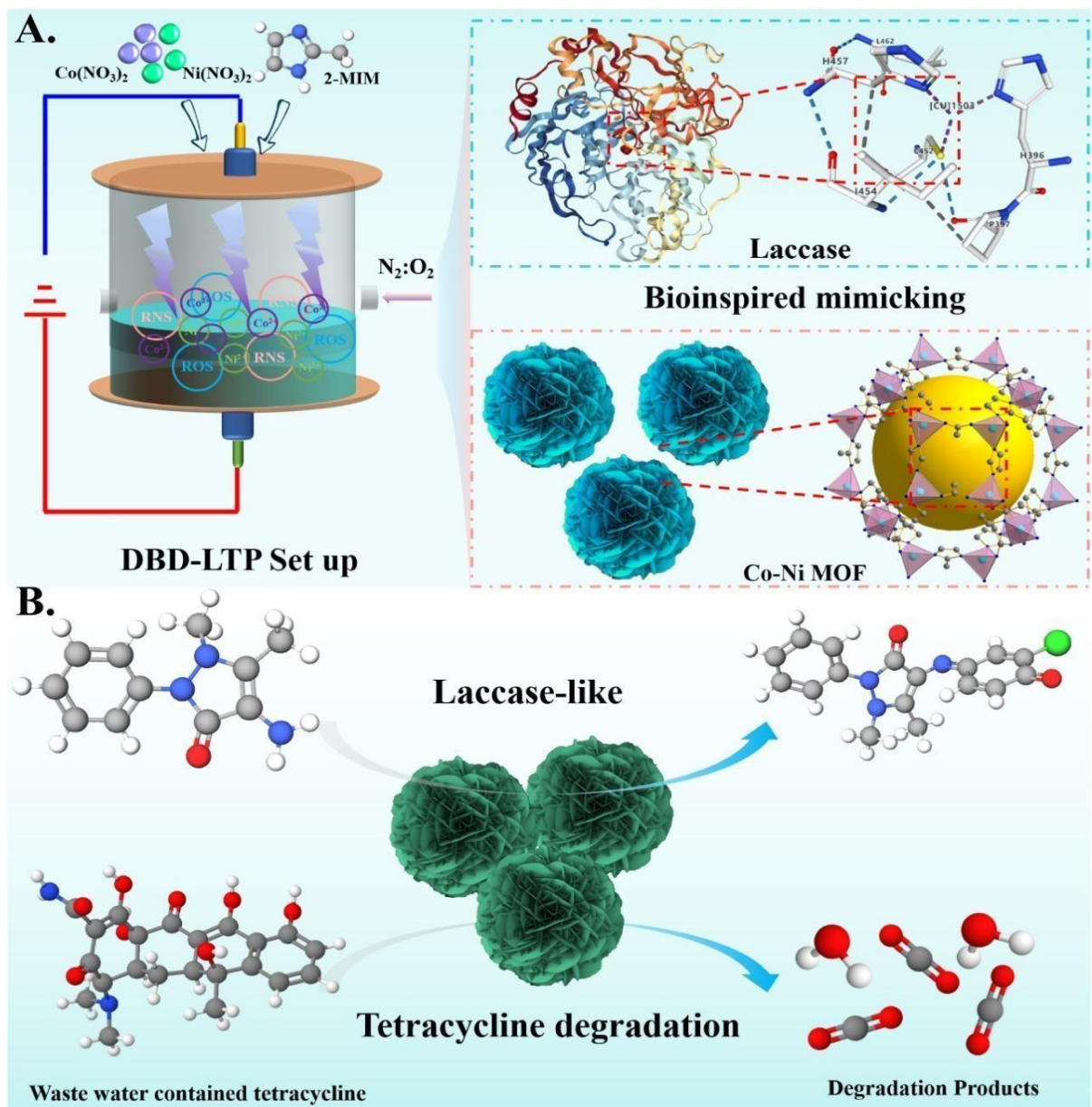

Scheme 1. Schematic illustration of A. the biomimetic design of MOFs and dielectric barrier discharge (DBD) form of low-temperature plasma (LTP) preparation process of CoNi-MOF; and B. the treatment of tetracycline in wastewater using the laccase-like activity of CoNi-MOF.

## 1. Materials and methods

### 1.1 Chemicals and apparatus

Co(NO)3·6H2O, Ni(NO)3·6H2O, methanol, ethanol, were purchased from Sinopharm Chemical Reagent Co., Ltd. Natural laccase we bought is S10189 from Yuanye company, whichis a crude enzyme and has activity more than 120 U/g. All reagents were analytical grade and used directly without further purification. The morphology was captured using a high-magnification transmission electron microscope (TEM, Thermo Fisher Talos F200s) and a field emission scanning electron microscope (SEM, ZEISS Gemini 450). UV–vis absorption was collected by UV–vis spectrophotometer (UV–Vis, UV-2550, Shimadzu) and an X-ray photoelectron spectrometer (XPS, Thermo Scientific ESCALAB Xi+) were utilized to collect and XPS spectra. More chemical information and analytical methods were provided in Supporting Information.

## 2.2 Preparation of the nanozymes using LTP technique

To obtain the CoNi-MOF nanozymes with different concentrations of oxygen vacancies, we changed the gas ratio in the plasma treatment. The specific procedure was as follows. First, Co(NO3)2 (10 mg/mL) and Ni(NO3)2 (10 mg/mL) were mixed with 2-Methylimidazole (50 mg/mL) ligand at a volume of 50 mL on a quartz reaction chamber. Then, different gases (pure N2, N2:O2=4:1, 1:1 and 1:4, pure O2) were introduced to the chamber for DBD-plasma treatment. The mixture was processed with a breakdown voltage of 55 kV for 1 minute with different gases. After the reaction, the yellow-green precipitate was washed three times with deionized water and ethanol, followed by freeze-drying.

## 2.3 Evaluation of the catalytic activities of CoNi-MOF

To verify the laccase-like activity of CoNi-MOF, CoNi-MOF (0.1 mg/mL) or laccase (0.5 mg/mL), 2,4-DP (1 mg/mL) and 4-AP (2 mg/mL, maintaining in excess) were added to an acetate-sodium acetate buffer solution (pH = 6.0, 0.1 M). The mixtures were left to stand at room temperature for 5 minutes before UV–Vis spectra at 400–700 nm was recorded.

## 2.4 Kinetic parameters determination

To determine the initial reaction rate, various concentrations of 2,4-DP (0.01, 0.1, 0.2, 0.3, 0.5, 1, 2, 5 and 10 mM) were reacted with 4-AP (2 mg/mL) and catalyzed by either CoNi-MOF nanozyme or laccase (0.1 mg/mL). The concentration of the product was calculated based on the Lambert–Beer Law, A=ε · b · c (ε is the molar absorbance coefficient 13.6 mM-1cm−1) [18]. The kinetic parameters were calculated based on the Michaelis-Menten equation:

$$V = V_{max}[S]/(K_m + [S]),$$

where V is the reaction rate, $V_{max}$ is the maximum reaction rate, [S] is the substrate concentration, and $K_m$ is the Michaelis constant. $K_m$ and $V_{max}$ were obtained from the Lineweaver–Burk plot: $1/V = K_m/V_{max} \times 1/[S] + 1/V_{max}$

## 2.5 Experimental procedures for antibiotics treatment

The tetracycline degradation experiment was carried out in a 200 mL conical flask. First, the CoNi-MOF catalyst (0.1 g/L) was dispersed in a tetracycline solution (50 mg/L, 50 mL). Before the reaction, the resulting mixture was stirred in the dark for 30 minutes to ensure the equilibrium of adsorption and desorption. At the same time, to explore the effect of different aeration conditions on the degradation efficiency of tetracycline, an aeration tube filled with air or pure $O_2$ (0.5 L/min) was submerged in the aqueous medium after the reaction started. 1 mL of the reaction suspension was taken at predetermined time intervals and filtered through a 0.22 μm nitrocellulose membrane. Methanol was used as a quencher in the reaction. The treated samples were analyzed using a high-performance liquid chromatograph (HPLC, Shimadzu): 10 μL of the sample was injected into an HPLC system equipped with an Inertsil C18 column (4.6 ×250 mm, 5 μm). The eluent consisted of 0.1% formic acid and acetonitrile (volume ratio of 20:80), the flow rate was 1.0 mL min-1, and the wavelength of the UV detector was set to 355 nm. The degradation rate of tetracycline was calculated based on the following formula:

$$\text{Degradation efficiency} = C/C_0,$$

where C is the concentration of tetracycline after treatment for different time periods, and $C_0$ is the initial concentration of tetracycline [19].

## Results and discussion

### 1.2 Synthesis and characterization of the CoNi-MOF

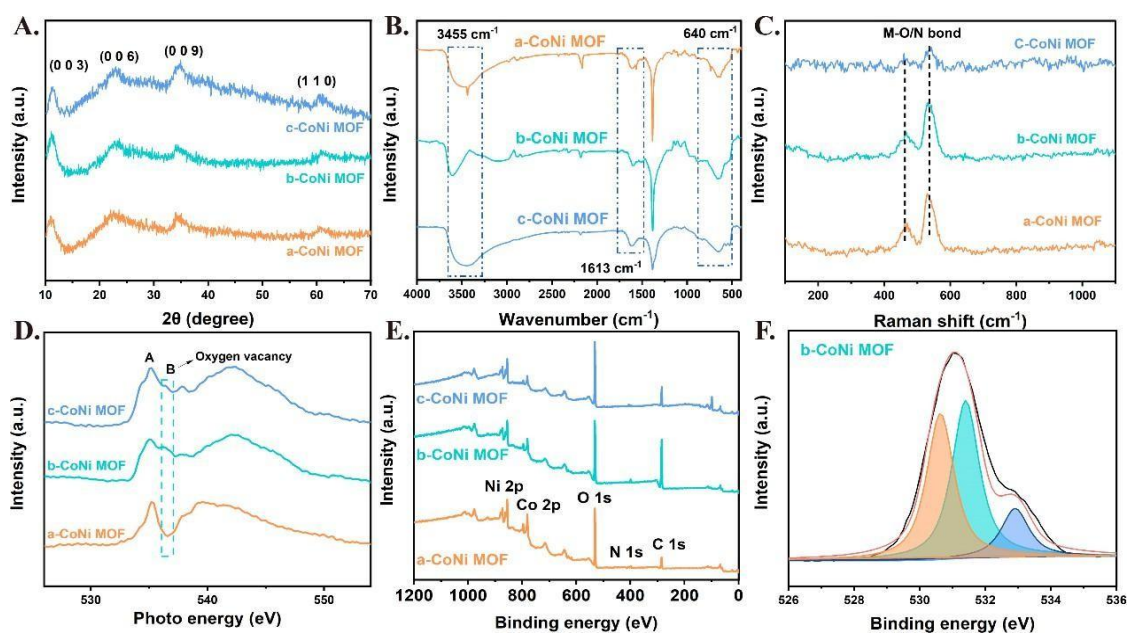

Figure 1. A. XRD, B. FTIR and C. Raman spectrum of CoNi-MOF; D. XAS results and E.

XPS survey of CoNi-MOF; F. O 1s spectrum of b-CoNi-MOF.

During the DBD treatment, a high concentration of reactive species was produced. To distinguish samples with different treatment parameters, the samples obtained by plasma treatment with N2:O2 ratios of 1:4, 1:1, and 4:1 were labeled as a-CoNi-MOF, b-CoNi-MOF, and c-CoNi-MOF, respectively. First, to confirm the successful synthesis of CoNi-MOF, powder X-ray diffraction (XRD) analysis was performed on three CoNi-MOF samples prepared under different gas treatment conditions. The results in Figure 1A showed that, with the assistance of plasma, $Ni^{2+}$ and $Co^{2+}$ successfully coordinated with the ligand 2-methylimidazole to form CoNi-MOF. The XRD patterns of all samples exhibited characteristic diffraction peaks at 2θ = 11°, 24°, 34°, and 61°, corresponding to the (0 0 3), (0 0 6), (0 0 9), and (1 1 0) crystal planes, respectively. This indicates the presence of a typical layered structure in the product, consistent with previously reported results, confirming the successful synthesis of CoNi-MOF [20]. Furthermore, FTIR was used to further validate the formation of CoNi-MOF. A new broad absorption band was observed at approximately 640 cm-1 in Figure 1B and Figure S1, attributed to the stretching vibrations of Ni-O and Co-O [21]. Additionally, broad peaks at 1613 cm-1 and 3455 cm-1 were assigned to the bending and stretching vibrations of hydroxyl groups from water molecules intercalated in the CoNi-MOF layers, respectively [22] . Figure 1C of Raman results also supported these findings, with peaks at 463 cm-1 and 520 cm-1 corresponding to the M-O/N (M=Co and Ni) stretching modes in CoNi-MOF, indicating the interaction between metal ion and oxygen/nitrogen atoms. These results collectively confirm the successful synthesis of CoNi-MOF.

Next, to investigate the effects of different plasma treatment conditions on CoNi-MOF, the reactive species generated under various gas parameters during the plasma preparation process were analyzed, and the unique structure of CoNi-MOF was verified using multiple characterization techniques. First, the results in Figure S2 showed that as the proportion of N2 increased during the treatment process, the plasma generated a large amount of reactive nitrogen species (RNS, such as NO3-) in the solution, while the amount of reactive oxygen species (ROS, such as ·OH) decreased. This change promoted the formation of the layered structure of CoNi-MOF and led to the generation of a significant number of oxygen vacancies. The result of O K-edge X-ray absorption spectroscopy (XAS) analysis is shown in Figure 1D. It is evident that the intensity of peak B (located at ~536.8 eV, attributed to oxygen vacancies) shows a significant increase compared to that of peak A (centered at ~535.1 eV) [23]. The result by X-ray photoelectron spectroscopy (XPS) in Figure 1F-G further confirmed the presence of Co, Ni, and O in CoNi-MOF. The O 1s spectrum could be deconvoluted into three characteristic peaks: oxygen atoms bonded to metals (530.3 eV), oxygen vacancy defects (531.6 eV), and surface-adsorbed H2O molecules (533.0 eV) [24,25]. Notably, oxygen vacancy defects began to appear when the N2:O2 ratio was 1:1 during plasma treatment, and the peak intensity at 531.6 eV was significantly higher when the N2:O2 ratio increased to 4:1. Based on calculations, the area ratios of oxygen vacancy defects were approximately 31% and 49.7%, respectively, with the latter being about 1.7 times higher than the former (Figure S3). This indicates that more oxygen defects appeared on the material surface as the proportion of N2 increased, consistent with the above test results. Therefore, XPS results, combined with the above analysis, provide critical evidence for the presence of a significant number of oxygen vacancies in CoNi-MOF after plasma treatment.

Subsequently, the FESEM and TEM images clearly show the morphological changes of the precursor during the plasma treatment process. When the plasma reaction atmosphere ratio was altered, the surface of CoNi-MOF was covered with a great number of thick, intertwined ultra-thin nanosheets. Notably, the sample exhibited a flower-like structure with nanosheets as the outer shell (Figure S4-7). This unique hollow structure significantly increased the specific surface area and provided abundant pores, which facilitates the rapid diffusion of the substrate. Additionally, the TEM and HRTEM images in Figures 2A-C confirmed the lattice fringe spacing of CoNi-MOF, corresponding to the (110) and (012) crystal planes of CoNi-MOF. Figure 2D-H showed the elemental mapping of the CoNi-MOF sample. It was observed that Ni, Co, and O elements are evenly distributed throughout the entire hollow octahedral structure, further confirming the successful synthesis of CoNi-MOF.

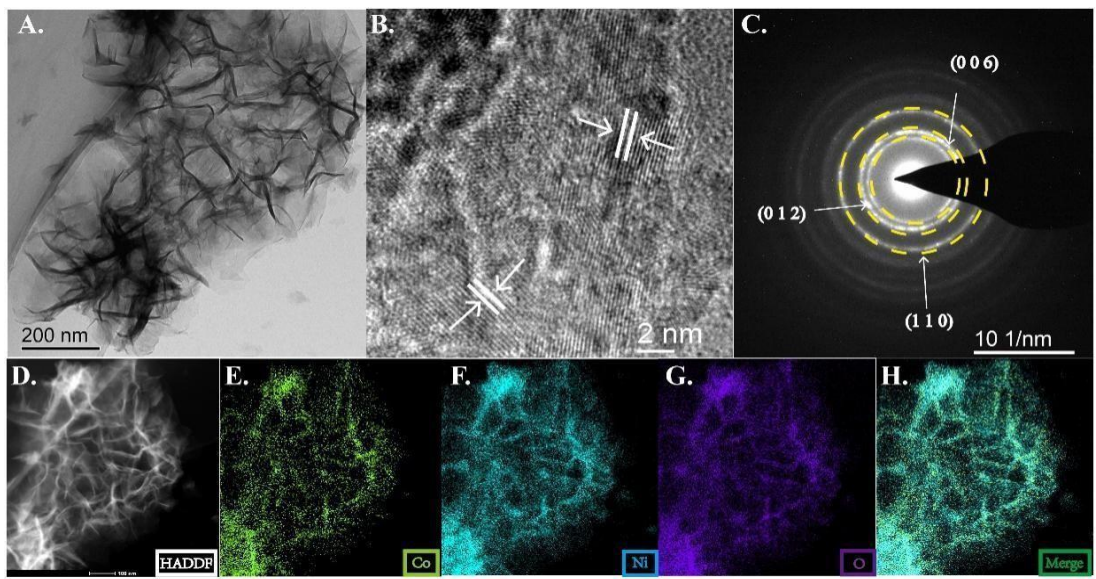

Figure 2. A. TEM images of and B. HRTEM image of b-CoNi-MOF; C. corresponding SAED images of b-CoNi-MOF; D-H. EDX elemental mappings of b-CoNi-MOF (Co, Ni, O and Merge images, Scale bar = 100 nm).

To investigate the chemical composition and element valence states of the three CoNi- MOFs, the XPS results are shown in Figure 3A, D and Figure S8. The full XPS spectrum indicates the presence of the major elements Ni, Co, and O. The main peaks at 780.9 and 796.4 eV correspond to Co 2p3/2 and Co 2p1/2, respectively. Additionally, the coexistence of Co2+ and Co3+ is clearly observed, with binding energies at 781.3 and 796.5 eV for Co2+ and at 780.2 and 795.8 eV for Co3+, suggesting the coexistence of Co2+ and Co3+, and indicating that some Co2+ are oxidized to Co3+ during the formation of CoNi-MOF [26]. The high-resolution Ni 2p XPS spectrum shows peaks at 873.0 and 855.3 eV corresponding to Ni 2p1/2 and Ni 2p3/2, respectively (Figure 3C), accompanied by two satellite peaks. Mixed valence states of Ni2+ (binding energies at 854.7 and 872.3 eV) and Ni3+ (binding energies at 855.6 and 873.2 eV) are also observed [27].

To gain deeper insights into the effect of plasma treatment on the composition and structure of CoNi-MOF, comprehensive X-ray absorption fine spectroscopy (XAFS) investigations were conducted, encompassing both XANES and EXAFS measurements. These advanced characterization techniques were employed to elucidate the alterations in electronic configurations and local coordination geometries surrounding the cobalt and nickel metal centers within the MOF framework following plasma treatment. Figures 3B and E showed the XANES spectra of the Co K-edge and Ni K-edge, compared with metal foils and corresponding oxide. As higher oxidation states typically lead to higher binding energies, as widely reported in the literature. For CoNi-MOF, the energy absorption threshold of Co is higher than that of Co foil, CoO, and Co3O4, indicating that the Co species in CoNi-MOF are mainly in an ionic state (close to Co3+). At the same time, the increased intensity of the Co K-edge XANES spectrum is attributed to the increased octahedral character and the higher number of unoccupied d-states in the Co ions, possibly due to vacancy formation [28]. Similarly, the Ni K-edge XANES spectrum showed no significant difference compared to standards such as Ni(OH)2, suggesting that the oxidation states of Ni are +2 and +3, maintaining an octahedral geometry. These results also suggest that Ni doping might promote the formation of oxygen vacancies. Figures 3C and F show the EXAFS analysis of the Co and Ni K-edge XAS spectra. The EXAFS fitting is shown in Figure S9-10. In the R-space FT-EXAFS spectrum of Co element in CoNi-MOF, a Co-O/N peak appears at 2.10 Å [29]. The EXAFS wavelet transform

(WT) analysis further provides the information on the coordination states of Co and Ni atoms in CoNi-MOF. The Co spectrum shows a clear Co-O/N bond formation, while the Ni spectrum shows a distinct Ni-O/N bond characteristic peak, which is similar to the structure of natural laccase (Figure 3J-G and Figure S11-12).

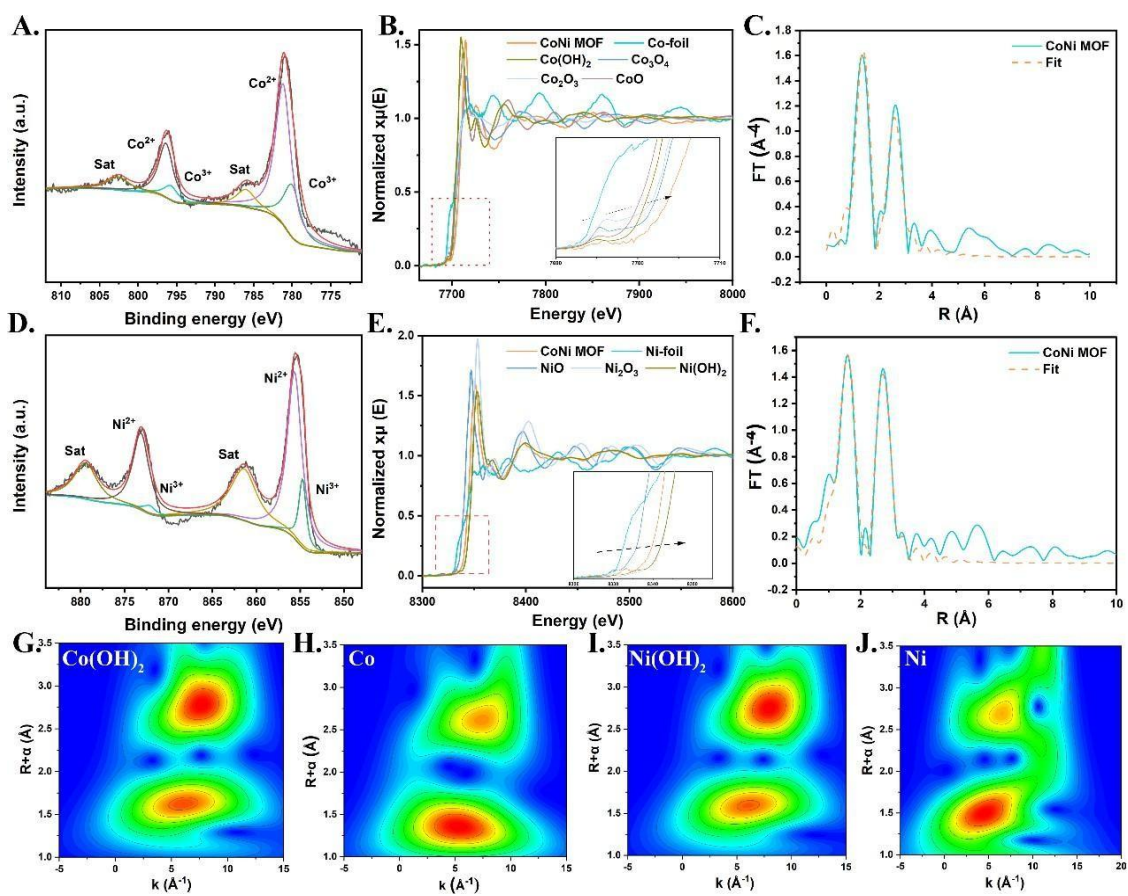

Figure 3. XPS analysis of the b-CoNi-MOF. A. Co 2p and D. Ni 2p; Ex situ near-edge X-ray absorption fine structure (NEXAFS) spectra for the oxygen K-edge of b-CoNi-MOF; X-ray absorption near-edge structure (XANES) spectra of the as-synthesized b-CoNi-MOF for B. and E; Least-squares EXAFS spectra of C. Co and F. Ni of b-CoNi-MOF in R space. G-J. Wavelet transform for the k2-weighted Co K-edge and Ni K-edge EXAFS signals of b-CoNi-MOF and the reference samples.

Based on the above characterization results, we can summarize the key role of plasma in the formation of CoNi-MOF and its possible mechanisms. First, plasma discharge generates a large amount of energy in a short time, facilitating the interaction and formation of new substances [30,31]. For example, plasma generates hydroxyl radicals at the gas-liquid interface, promoting the coordination of $Ni^{2+}$ and $Co^{2+}$ with 2-methylimidazole in the solution via increasing the Lewis acidity of the metal ions [32,33]. Simultaneously, other reactive species generated by plasma (such as h+ and e-) can accelerate the oxidation of active metals, generating more active sites on the CoNi-MOF surface and promoting the rapid growth of nanostructures. Second, plasma treatment may also affect the formation of oxygen vacancies on the CoNi-MOF. Specifically, increasing the nitrogen gas ratio during the plasma process can enhance the concentration of nitrate ions in the solution while reducing the concentration of hydroxyl radicals [34]. This promotes the formation of the MOF layered structure and results in the generation of numerous oxygen vacancies [35]. Therefore, compared to traditional hydrothermal co-precipitation methods, the time required for preparing nanozyme using DBD-LTP can be significantly shortened, without the need for precipitants or surfactants, and reactions can occur under ambient conditions. Moreover, the controllability of reaction conditions provides a wide range of possibilities for activity regulation, facilitating practical applications in various scenarios.

**1.3 Laccase-like activity of the CoNi-MOF nanozyme**

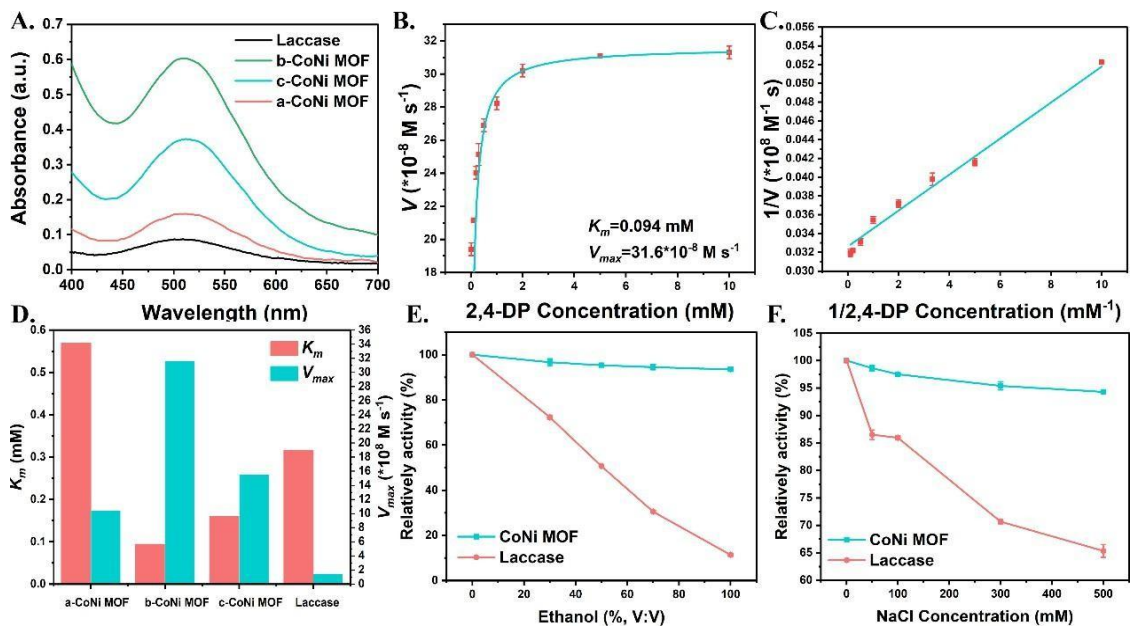

Figure 4. A. UV-vis absorption spectroscopy for the laccase-like activity of CoNi-MOF nanozymes.; B. Steady-state kinetics and C. Lineweaver-Burk plot of the reciprocals of initial rate vs 2,4-DP concentration of CoNi-MOF; D. Comparison of the $K_m$ and $V_{max}$ of different CoNi-MOF; Relative activity of CoNi-MOF nanozymes and laccase at varying E. ethanol content and F. NaCl concentration.

Due to the structural similarities between CoNi-MOF and natural laccase, the laccase-like activity of CoNi-MOF was evaluating using 2,4-dichlorophenol (2,4-DP) as the catalytical substrate and 4-aminoantipyrine (4-AP) as the colorimetric indicator. As shown in Figure 4A, both laccase and CoNi-MOF nanozyme catalyze the oxidation of 2,4-DP. The oxidized 2,4-DP subsequently reacted with 4-AP to form a distinct red-colored quinone-imine derivative, showing a characteristic UV absorption peak at approximately 510 nm. Further comparative studies were conducted to investigate the excellent laccase-like activity and the effects of pH and temperature on the laccase-like activity of CoNi-MOF nanoparticles (Figures S13). The CoNi-MOF laccase-like nanozyme demonstrated better laccase-like activity in a nearly neutral environment (pH=6.0), with the highest catalytic efficiency observed at 37 °C, indicating its excellent thermal stability.

To compare the catalytic activity of natural laccase and CoNi-MOF laccase-like nanozyme with different oxygen vacancy concentrations, the initial reaction rates at different substrate concentrations were measured, and the Michaelis-Menten model and Lineweaver-Burk fitting were used to calculate the $K_m$ and $V_{max}$ values for three CoNi-MOF laccase-like nanozyme with different oxygen vacancy concentrations. As shown in Figure 4B-C and Figure S14, compared with natural laccase, b-CoNi MOF laccase-like nanozyme had a lower $K_m$ value of 0.094 mM and a higher $V_{max}$ of $31.6 \times 10^{-8}$ M s-1, indicating a better substrate binding affinity and reaction velocity during the catalytic oxidation of 2,4-DP. Furthermore, compared to other nanozymes with laccase-like activity reported in previous literature, b-CoNi MOF laccase-like nanozyme exhibited the highest kinetics (Figure 4D and Table S2). The above results also show that the appropriate oxygen vacancy concentration has advantages in regulating laccase activity. Thus, based on the comparative analysis of enzyme-like activities among a-CoNi MOF, b-CoNi MOF, and c-CoNi MOF, we confirmed that b-CoNi MOF demonstrated superior catalytic performance. Consequently, b-CoNi MOF was selected for further experiments to explore its practical applications.

Additionally, the effect of interference chemical reagents on the catalytic activity of laccase-like nanozyme and natural laccase was investigated. Both CoNi-MOF laccase-like nanozyme and laccase were incubated at varying concentrations for 60 minutes. As illustrated in Figure 4E, a concentration-dependent inhibition of catalytic performance was observed for both CoNi-MOF laccase-like nanozyme and laccase. Notably, at 70% ethanol concentration, laccase almost completely lost its activity, while CoNi-MOF laccase-like nanozyme maintained over 93% activity. Since laccase is commonly used for industrial wastewater treatment, the impact of ionic strength on its activity was also studied. The result in Figure 4F revealed that natural laccase exhibited substantial sensitivity to NaCl concentration. This phenomenon can be attributed to the disruption of the enzyme's electrostatic equilibrium and conformational integrity under high-salinity conditions. However, CoNi-MOF laccase-like nanozyme retained more than 95% of their activity even at a NaCl concentration of 500 mM.

**1.4 Catalytical mechanism for explaining the CoNi-MOF laccase-like activity**

Accurate investigation of the catalytic mechanism of CoNi-MOF is essential for further improving its catalytic performance. In this section, we used b-CoNi-MOF as an example to explore the potential catalytic mechanism of CoNi-MOF in its laccase-like activity. First, the result by electrochemical impedance spectroscopy (EIS) revealed that CoNi-MOF exhibited higher electron mobility (Figure 5A). Also, it was demonstrated that the catalytic process of CoNi MOF was similar to the same 4e- process for producing $H_2O$ as natural laccase (Figure S15). Additionally, the catalytic activity of CoNi-MOF was significantly suppressed under nitrogen ($N_2$) conditions but increased dramatically under oxygen ($O_2$) conditions (Figure 5B), verifying the $O_2$-dependent laccase-like catalytic oxidation of CoNi-MOF.

Next, free radicals generated by CoNi-MOF catalyzing $O_2$ were identified using various methods to clarify the reactive intermediates produced during the catalytic process and determine the catalytic pathway. We investigated the effects of different free radical scavengers (such as isopropanol and D-mannitol (·OH scavengers), 1,4-Benzoquinone and Superoxide dismutase ($O_2·^-$ scavengers), and L-tryptophan and L-histidine ($^1O_2$ scavengers)) on the activity of CoNi MOF nanozyme [36,37]. As the concentration of ·OH scavengers and $^1O_2$ scavengers increased, the laccase-like activity gradually weakened (Figure 5C). Subsequently, to elucidate the reactive oxygen species generated during the catalytic process, electron spin resonance (ESR) spectroscopy was employed to characterize the transient radical intermediates produced by the CoNi-MOF system, with the result presented in Figure 5D. The spin trapping technique utilizing 5,5-dimethyl-1-pyrroline N-oxide (DMPO) as the trapping agent revealed the formation of hydroxyl radicals (·OH), evidenced by the characteristic DMPO/·OH adduct signature in the ESR spectrum. This adduct exhibited a distinctive quartet signal with relative peak intensities of 1:2:2:1, confirming the generation of •OH radicals during the catalytic reaction. Similarly, the combination of 2,2,6,6-Tetramethyl-4-piperidone hydrochloride (TEMP) and $^1O_2$ produced an ESR signal.

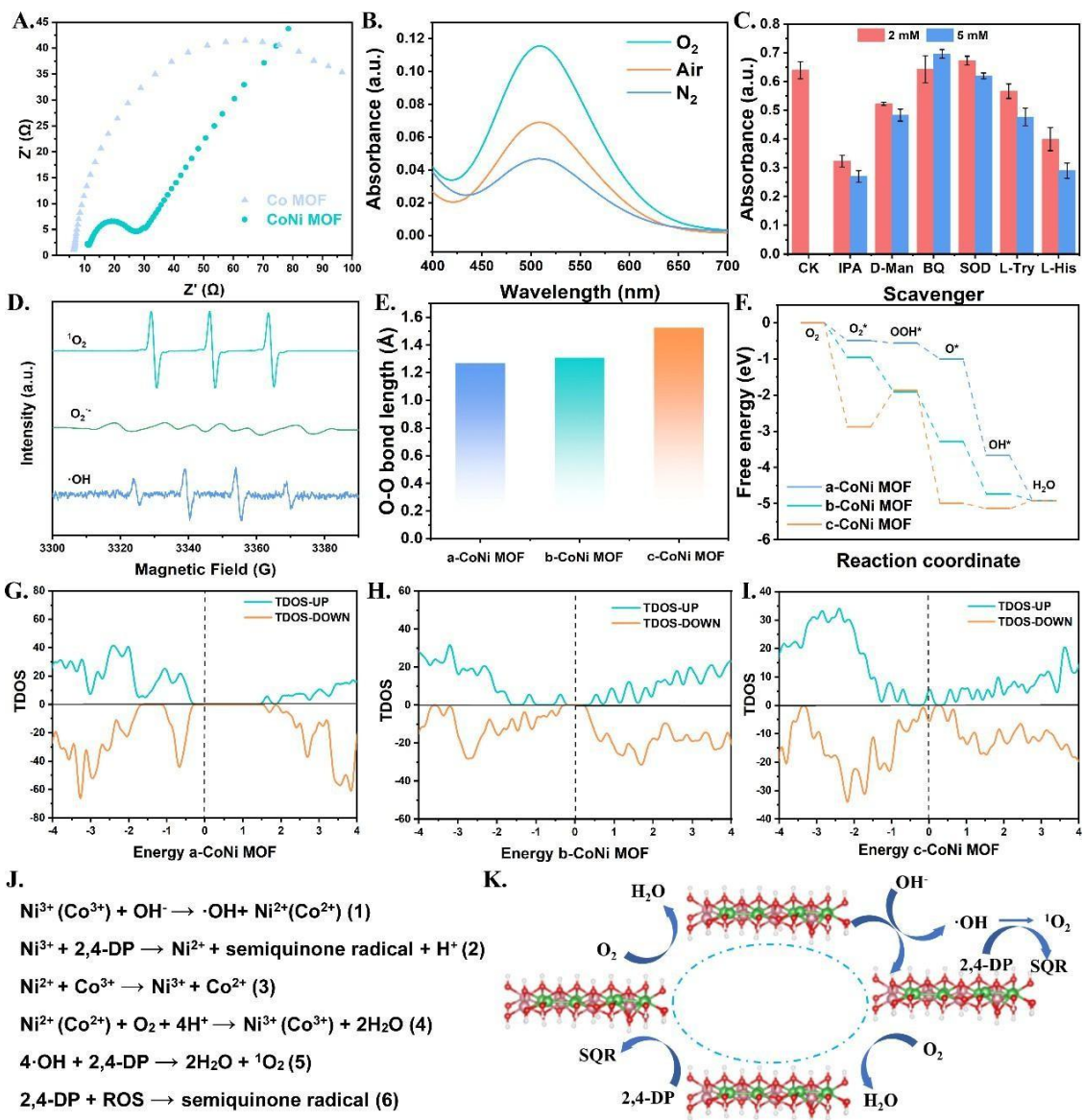

Figure 5. A. EIS spectra of CoNi-MOF and Co-MOF; B. UV-vis absorption spectra of the catalytic oxidation of 2,4-DP by CoNi-MOF under different atmospheres; C. Effect of free radical scavengers on the performance of CoNi-MOF; D. ESR spectra of different systems; E. Comparison towards O-O the bond length of CoNi-MOF; F. Changes in free energy along the reaction pathways; G-I. Total density of states (TDOS) of CoNi-MOF; J-K. LAC-like catalytic mechanism of CoNi-MOF under neutral condition.

To further elucidate the theoretical possibilities behind CoNi-MOF's excellent nanozyme activity, density functional theory (DFT) calculations were performed to analyze the substrate adsorption and reaction pathway. The study examined the types and pathways of O2 conversion to active species on the CoNi-MOF surface. Firstly, Figure 5E displays the O-O bond length on different CoNi-MOFs. O2 adsorbs at the active site, and the O-O bond length of the oxygen molecule stretches from 1.27 to 1.53 Å, making it more easily activated [38]. Figures 5F illustrates the O2 adsorption and desorption process at the CoNi-MOF interface, and the Gibbs free energy (ΔG) for each step was calculated. The ΔG values represent the Gibbs free energy for each step, with the highest ΔG step being the rate-limiting step. Comparing these values, it is observed that the formation of oxygen vacancies enhances the intrinsic activity at the interface, improving O2 adsorption, making the transformation of O2 into OOH easier, and shifting the rate-limiting step to the OH + H dehydration step with an energy of -0.192 eV. This makes the reaction more favorable with a more balanced energy profile [39]. However, when vacancies increase further, more metal sites are exposed on the surface, which enhances adsorption but makes desorption more difficult. Therefore, a moderate vacancy concentration is optimal for favorable reactions. This also further explains why b-CoNi-MOF exhibits stronger enzyme-like catalytic activity. Additionally, the density of states (DOS) results in Figure 5G-I and Figure S16 showed that after introducing oxygen vacancies, the energy bands of CoNi-MOF shift to higher energy regions and broaden, strengthening the hybridization along the Co 3d-O 2p-Ni 3d pathway and stimulating charge transfer, thus enhancing the metallic properties of CoNi-MOF.

Based on these results, we therefore hypothesize that the catalytic reaction may follow a bimetallic synergistic catalysis mechanism (Figure 5J-K). In a neutral environment, Co can also serve as an active site, enhancing the laccase-like activity of CoNi-MOF (1). $Ni^{3+}$ gains an electron from 2,4-DP on the CoNi-MOF surface and is reduced to $Ni^{2+}$, resulting in the oxidation of 2,4-DP to a semiquinone radical (SQR) (2). Then, $Ni^{2+}$ is oxidized to $Ni^{3+}$ by $Co^{3+}$ (3). Simultaneously, $Ni^{2+}$ and $Co^{2+}$ can be oxidized by O2 to $Ni^{3+}$ and $Co^{3+}$ (4). The resulting ·OH radical polarizes to generate $^1O_2$ (5). Finally, all generated reactive oxygen species (ROS) (·OH and $^1O_2$) oxidize 2,4-DP to SQR (6) [40].

### 3.4. Reactive oxygen species generation and tetracycline degradation

Considering the catalytic performance of CoNi-MOF, which is comparable to that of laccase and capables of utilizing oxygen to oxidize various aromatic substrates. In this study, we focused on b-CoNi-MOF and investigated its ability to remove tetracycline, based on the above laccase-like activity verification. As shown in Figure 6A and Figure S17-19, CoNi- MOF achieved an 90% tetracycline degradation efficiency over a 70-minute treatment period and showed excellent anti-interference ability, which aligns with the trends observed in its laccase-like characteristics. At the same time, CoNi-MOF also showed excellent degradation ability for tetracycline antibiotics such as oxytetracycline and chlortetracycline, indicating the universal significance of CoNi MOF for pollutant degradation (Figure S20). Notably, based on the results that CoNi-MOF exhibited laccase-like activity through the activation of O2, we therefore hypothesized that increasing the oxygen content could enhance the accumulation of ROS, thus promoting tetracycline degradation.

To test this hypothesis, we first measured the changes in dissolved oxygen (DO) concentration in water after exposure to different air/oxygen conditions. As shown in Figure S21A, the DO concentration in the system increased over time. Next, we examined the exact amount of ROS generated in the optimal CoNi-MOF catalytic system. When air was introduced, ROS such as ·OH and $^1O_2$ were generated at moderate DO concentrations. However, when pure oxygen was introduced, the ROS production increased significantly, and this was confirmed by EPR results (Figure 21 B-E). The noticeable differences in ROS production under different gas environments indicate that the ROS production was primarily originated from O2, suggesting that O2 activation mainly occurred at the gas-bubble-catalyst interface.

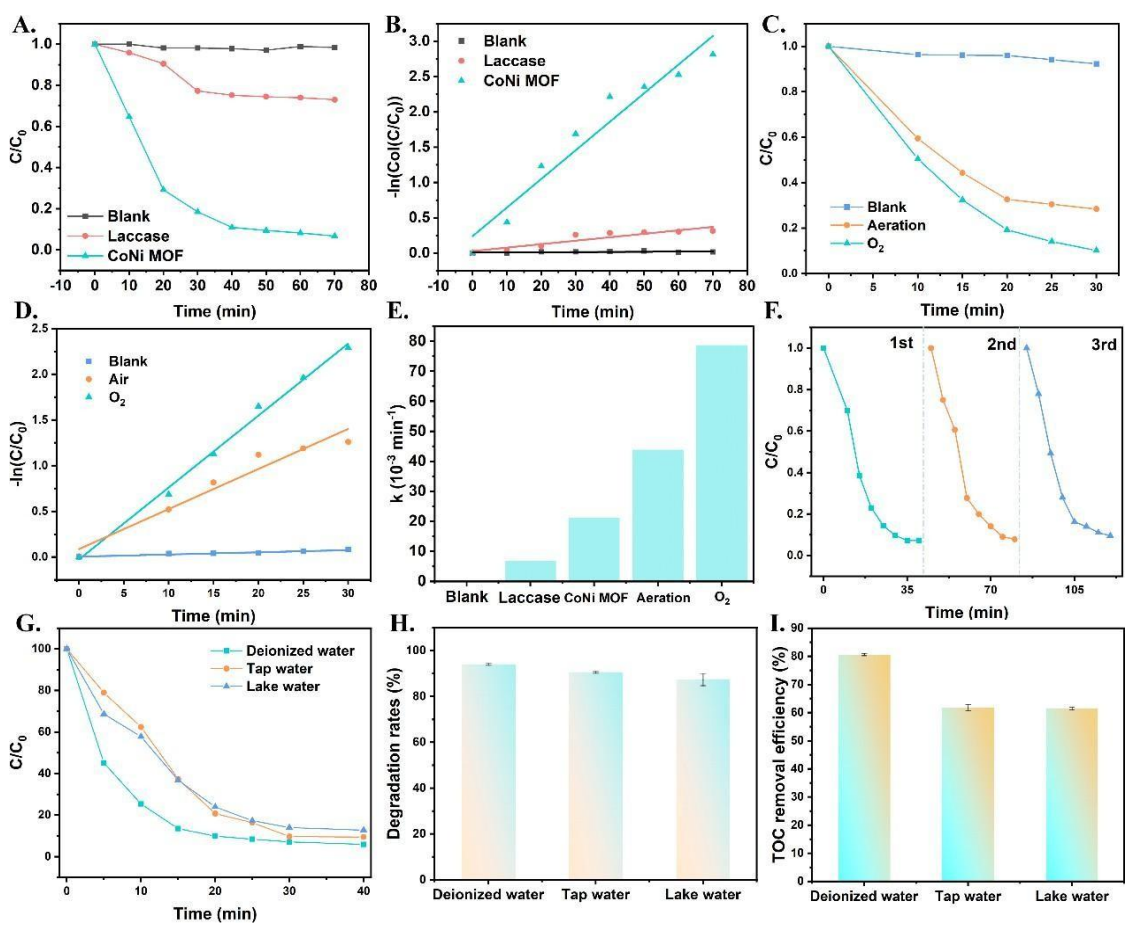

Figure 6. A. Degradation of tetracycline using laccase and CoNi-MOF nanozymes (tetracycline is 20 mg/L, catalysts are 100 mg/L, 25°C, pH=6.0, 3 replicates). B. The pseudo-first-order

kinetic constant fits the degradation kinetic curves; C. Degradation rate and D. the pseudo-first-order kinetic constants with different atmosphere conditions (tetracycline is 50 mg/L, the flow rate of O2 is 0.5 L/min.); E. Comparison of kinetic constants with different reaction conditions; F. Cycling stability of CoNi-MOF in degrading tetracycline; G-H. Degradation of tetracycline and I. TOC removal rates in wastewater matrices (Deionized water, Tap water and Lake water).

Therefore, we introduced aeration or pure O2 into the tetracycline degradation system. The results showed that application of CoNi-MOF could more efficiently treat the higher concentrations of tetracycline, as compared to many other reported methods (Table S3-4). The reaction rate constants increased by 2.06-fold and 3.69-fold under aeration and oxygen-enriched conditions, respectively, compared to the control (Figures 6C-E), demonstrating that CoNi-MOF has good potential for integration with aeration systems in wastewater treatment. These values were derived from kinetic studies and reflected the acceleration of the catalytic reaction. Importantly, to assess the stability and reusability of CoNi-MOF, repeated cycling degradation experiments of tetracycline showed that the degradation efficiency remained nearly unchanged after three cycles, confirming the high stability and reusability of CoNi-MOF in tetracycline degradation (Figure 6F).

Furthermore, to demonstrate the feasibility of application, a small-scale circulating water treatment device was set up in the laboratory to simulate an actual wastewater treatment process (Figure S22). The setup includes a circulating reactor, a peristaltic pump, and a filtration membrane. The wastewater is circulated repeatedly through the membrane containing the catalyst under the action of the peristaltic pump. Figures 6G-H showed the treatment results for real contaminated tap water and lake water containing antibiotics. When the CoNi-MOF system was used to treat antibiotic-contaminated deionized water, tap water and lake water, satisfactory purification effects were achieved (94.11, 90.58% and 87.23%, respectively). Total Organic Carbon (TOC) is an important indicator for evaluating organic pollutants in wastewater samples and is commonly used to assess the quality of treated water. After 30 minutes of continuous treatment, the TOC degradation efficiency of simulated antibiotic wastewater was 80.59%, 61.78%, and 61.44% (Figure 6I), far exceeding the drinking water quality standards set by the national and U.S. Environmental Protection Agency (EPA). In addition, the amount of $Co^{2+}$ and $Ni^{2+}$ among supernatant after degradation was measured by ICP-MS, the leaching amount of $Co^{2+}$ is 17.96 μg/L while $Ni^{2+}$ is 62.26 μg/L, indicating the environmental safety of the CoNi-MOF.

**3.5 Possible degradation pathway and toxicity of the products**

To investigate the degradation process and the degree of mineralization of tetracycline, we initially employed 3D EEM fluorescence spectroscopy for analysis. The results shown in Figure S23 revealed that due to the multiple electrons of the electron-withdrawing group of tetracycline, no obvious fluorescence signal peak was detected in the tetracycline solution at the beginning of the reaction. However, as the reaction continued, an obvious peak was detected in the solution at Ex/Em= (325-375 nm)/(375-450 nm), corresponding to the humic acid fluorescence region. Especially after 30 minutes, it showed that tetracycline had decomposed and produced a variety of low-molecular intermediates. As the photocatalytic time was further extended, the fluorescence intensity continued to change, indicating that tetracycline was further decomposed. These results support that tetracycline is gradually transformed into small organic products, which is aligns to the results of LC-MS analysis later. Thus, high-performance liquid chromatography-mass spectrometry (HPLS-MS) was used to detect and analyze its degradation pathway and the degradation mechanism.

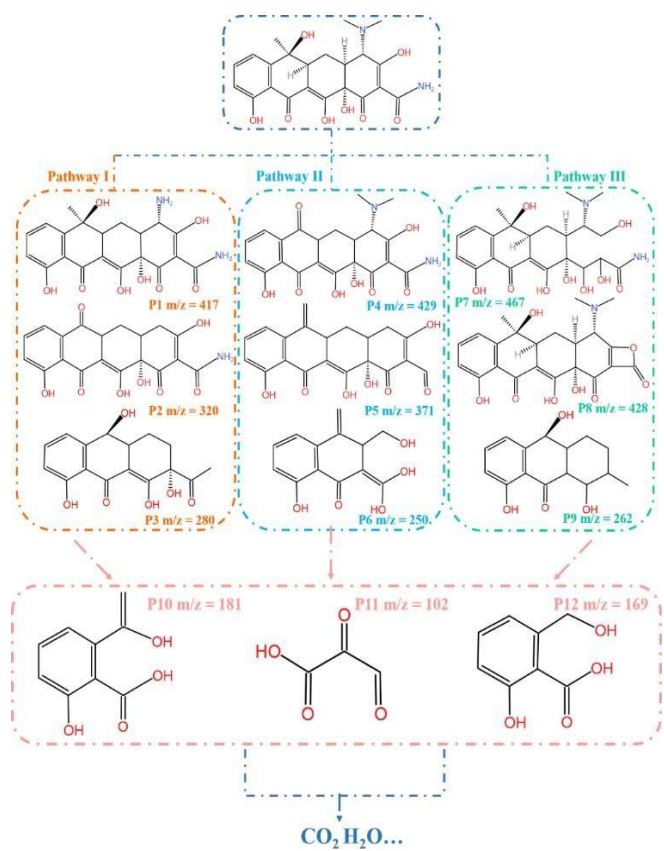

Figure 7. Schematic diagram of possible tetracycline degradation pathways by CoNi-MOF laccase-like nanozyme.

The Fukui function (CFF) provides reasonable speculation on the degradation of tetracycline firstly, which helps to judge the reliability of our hypothesized LC-MS-based degradation process. The calculation results are shown in Figure S24 and results show that the double bonds, phenolic and amine groups in the tetracycline have high electron density and are easily attacked by active oxygen groups (·OH and $1O_2$), resulting in a series of intermediates. Combining the calculation results with the results of LC-MS data (Figure S25), the possible degradation pathways are shown in Figure 7, which comprehensively summarizes the various pathways involved in the degradation, indicating that tetracycline gradually degrades to lower molecular products: In the first degradation pathway, the tetracycline molecule was attacked by ·OH, underwent a demethylation reaction to generate P1 (m/z=417), and then was converted to P2 (m/z=320) through deamidation and hydroxylation [41]. The fourth ring was cleaved and the amino group was detached to generate fragment P3 (m/z=280). In addition, fragment P4 (m/z=427) came from the degradation of tetracycline through a dehydration reaction. P4 lost the amino group and methyl group due to the attack of $1O_2$ to form fragment P5 (m/z = 371). At the same time, under the attack of active free radicals, the third ring of P2 was cleaved and underwent hydroxylation to generate fragment P6 (m/z=250). The formation of P7 (m/z=467) was due to the reaction between ·OH and tetracycline (m/z=445), resulting in the addition of -OH groups. It was then converted to P8 (m/z=428) through deamidation and dihydroxylation. Subsequently, due to the low C-N bond energy, under the continuous attack of active free radicals, the carbon cation ring structure was oxidatively fragmented and converted to product P9 (m/z=262). These intermediates were further oxidized, resulting in the decomposition of the remaining ring structure to form products P10 (m/z = 181), P11 (m/z = 102), and P12 (m/z=169). Finally, product underwent mineralization and decomposed into $CO_2$, $H_2O$, and other inorganic substances [5,7].

The migration and transformation of antibiotic residues and their by-products may pose unknown ecological risks. Therefore, it is necessary to evaluate the potential toxicity of these tetracycline intermediate products in the CoNi-MOF system. The Toxicity Estimation Software Tool (T.E.S.T.) software was used to investigate four important indicators for tetracycline and its 12 intermediates, with the results shown in Figure 8. The lethal concentration for 50% of water fleas (LC50-48 h) for tetracycline was 8.7 mg/L, indicating that tetracycline is a "toxic" pollutant. However, most of the transformation products are compounds with relatively lower toxicity (Figure 8A). Notably, the LC50-48 h values for small molecule products, especially P11 and P12, increased significantly to 239.94 mg/L and 450.96 mg/L, respectively, suggesting that the toxicity decreased after degradation. Similarly, as shown in Figure 8B, the LC50 (96

h) for fish, tetracycline itself showed high toxicity, with an LC50 (96 h) value of 0.25 mg/L. All intermediates had LC50 (96 h) values were higher than that of tetracycline, indicating that the acute toxicity of the intermediates was lower. Furthermore, Figure 8C-D also showed that the developmental toxicity and mutagenicity of tetracycline surpassed 0.5, classifying it as a "developmental toxicant" and "mutagenic positive". Most intermediates had lower developmental toxicity and mutagenicity compared to tetracycline. After deep oxidation, especially for several small molecule transformation products, negative mutagenic results and "non-toxic developmental" effects were observed, indicating a substantial reduction in toxicity during the green degradation [42].

In addition, we also applied *Escherichia coli* and zebrafish models to verify the degradation effect and bio-safety of the treatment. The *E. coli* survival experiment in **Figure S26** was performed to explore the toxicity of tetracycline and its degradation products. The results showed that compared to the untreated group, the survival rate gradually increased with prolonged treatment time, indicating that tetracycline was successfully degraded. Similarly, the zebrafish assessment in **Figure S27** showed that, in contrast to the acute toxicity observed in the high-concentration antibiotic group, the survival rate of zebrafish was not affected by CoNi-MOF laccase-like nanozyme. Furthermore, the wastewater solution treated by the CoNi-MOF laccase-like nanozyme system almost did not cause any change in the zebrafish survival rate. These results confirm the high biocompatibility of green CoNi-MOF laccase-like nanozyme, suggesting its potential for significant practical application.

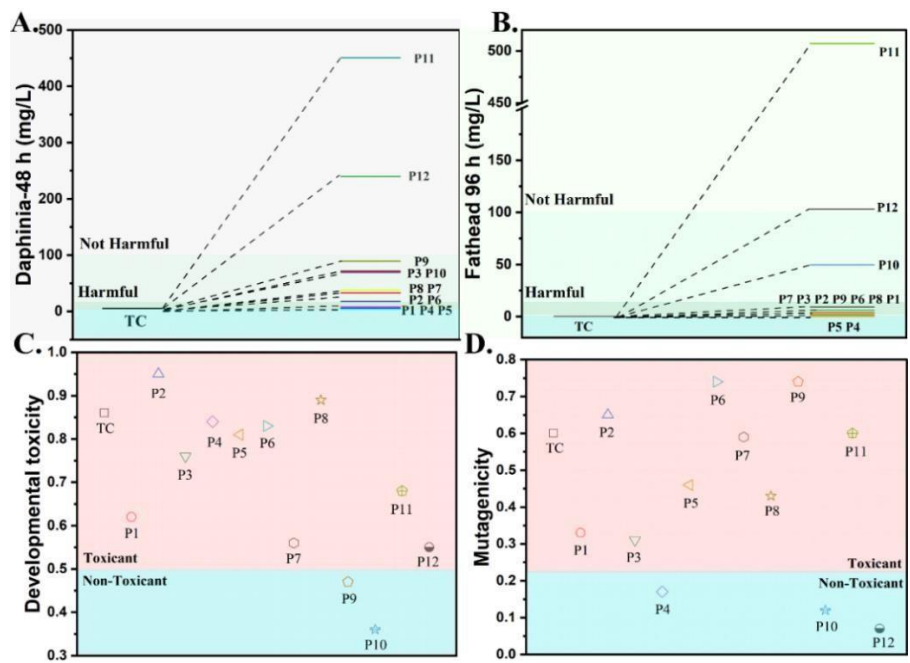

Figure 8. A. Daphinia magna LC50–48 h; B. fathead minnow LC50–96 h; C. developmental toxicity and D. mutagenicity of tetracycline and its intermediates in CoNi-MOF-aeration system.

## 4 Conclusions

In summary, we successfully prepared CoNi-MOF laccase-like nanozymes with different oxygen vacancy concentrations using DBD plasma, and showed that they could be used efficiently for the antibiotic pollution treatment. Compared to natural laccase, CoNi-MOF nanozymes demonstrate superior catalytic activity and greater stability under a wide range of challenging environmental conditions (extreme pH, high salt and high ethanol concentration conditions). Our experimental and theoretical results revealed that the oxygen vacancy played an important role in the laccase-like activity of CoNi-MOF.

Due to the excellent catalytic property, the as-prepared CoNi-MOF nanozymes could promote the oxidation of tetracycline, and notably, the involved aeration significantly increased the dissolved oxygen content, thereby enhancing the degradation of tetracycline. Within 30 minutes, a degradation efficiency of over 90% was be achieved. The results highlight the potential of CoNi-MOF as a sustainable and effective alternative to natural laccase in environmental remediation applications. Therefore, this study not only presents a simple and efficient approach for synthesizing high-performance laccase-like nanozymes based on low-temperature plasma technology, but also provides a promising and environmentally friendly strategy for addressing antibiotic pollution in the environment.

**Acknowledgment**

This work was partly supported by the National Natural Science Foundation of China (Grant No. 11635013 and No. 11775272), and Anhui Provincial Key Research and Development Plans (202004i07020014). The EXAFS experiments were performed at Shanghai